\documentclass[10pt,twocolumn,letterpaper]{article}
\usepackage{iccv}
\usepackage{times}
\usepackage{epsfig}
\usepackage{graphicx}
\usepackage{amsmath}
\usepackage{amssymb}
\usepackage{subfloat}
\usepackage{subfigure}
\usepackage{color}
\usepackage{multirow}

\usepackage{algorithm}
\usepackage[noend]{algpseudocode}

\usepackage[pagebackref=true,breaklinks=true,letterpaper=true,colorlinks,bookmarks=false]{hyperref}

\iccvfinalcopy 

\def\iccvPaperID{2106} 

\newcommand{\RED}{\textcolor{red}}
\newcommand{\BLUE}{\textcolor{blue}}
\ificcvfinal\fi
\begin{document}

\title{Automatic Content-aware Projection for 360$^\circ$ Videos}

 \author{Yeong Won Kim $^{1,}$\thanks{These authors contributed equally to this work.} $^{,}$\thanks{This work was done when the author was in Gwangju Institute of Science and Technology (GIST).} \and 
 Chang-Ryeol Lee$^{2,}$\footnotemark[1] \and
 Dae-Yong Cho$^2$ \and
 Yong Hoon Kwon$^2$ \and  
 Hyeok-Jae Choi$^2$  \and  \vspace{1mm}
 Kuk-Jin Yoon$^2$ \and   
 $^1$VUNO, South Korea \\ 
 {\tt\small ywon.kim@vuno.co} \\
 $^2$Gwangju Institute of Science and Technology (GIST), South Korea\\   
 {\tt\small \{crlee, dycho, hyeokjae94, yhkwon, kjyoon\}@gist.ac.kr} \\
 }

\maketitle
\thispagestyle{empty}

\begin{abstract}
To watch 360$^\circ$ videos on normal 2D displays, we need to project the selected part of the 360$^\circ$ image onto the 2D display plane.
%
In this paper, we propose a fully-automated framework for generating content-aware 2D normal-view perspective videos from 360$^\circ$ videos. Especially, we focus on the projection step preserving important image contents and reducing image distortion.  
%
Basically, our projection method is based on Pannini projection model. 
At first, the salient contents such as linear structures and salient regions in the image are preserved by optimizing the single Panini projection model.
Then, the multiple Panini projection models at salient regions are interpolated to suppress image distortion globally.
Finally, the temporal consistency for image projection is enforced for producing temporally stable normal-view videos.
%
%
Our proposed projection method does not require any user-interaction and is much faster than previous content-preserving methods. 
It can be applied to not only images but also videos taking the temporal consistency of projection into account.  
%
%
Experiments on various 360$^{\circ}$ videos show the superiority of the proposed projection method quantitatively and qualitatively. 
\end{abstract}

\begin{figure*}[t]
\centering 
\includegraphics[width=1.0\linewidth]{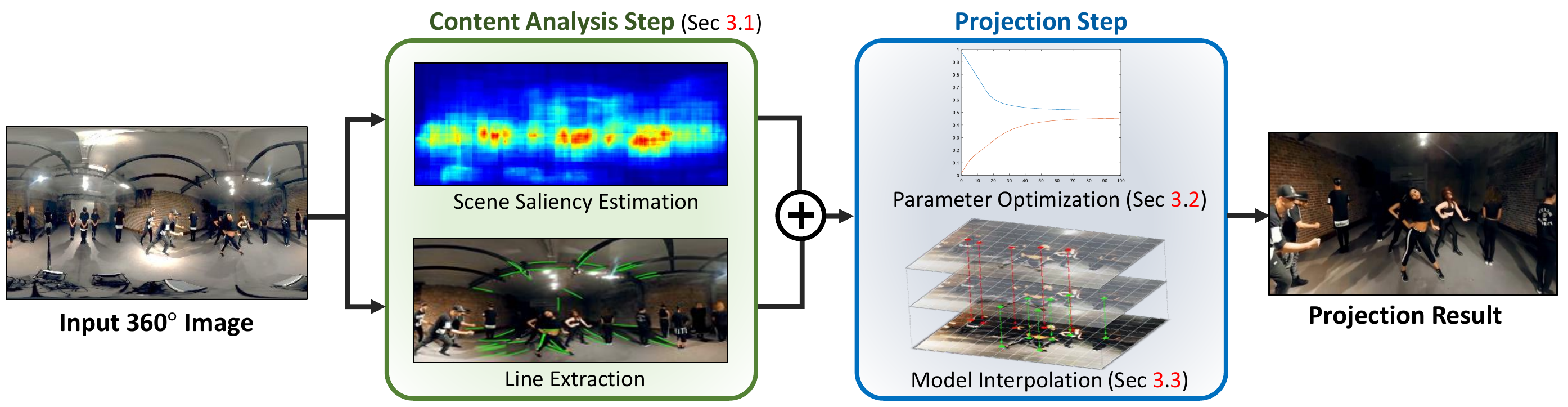} \vspace{-16pt}
\caption{\label{fig:framework} Overall framework for generating 2D normal-view videos from 360$^\circ$ videos. In this paper, although we cover the contents analysis step in Sec.~\ref{sec:content_analysis}, we mainly focus on the projection step. The contents analysis step such as viewpoint selection and salient region  extraction can be replaced with any other methods.}
\vspace{-7pt}
\end{figure*}

\vspace{-10pt}\section{Introduction}

Unlike traditional cameras which have a limited field of view (FOV), 360$^\circ$ cameras take omni-directional images at once.
Therefore, it becomes much easier to capture the objects of interest or meaningful events, whereas the traditional cameras require careful viewpoint control.
Recently, low-cost 360$^\circ$ cameras have been released thanks to the advance of Micro-Electro-Mechanical System (MEMS) technology, and lots of 360$^\circ$ videos are available on content distribution sites such as Youtube and Facebook.

To watch the 360$^\circ$ videos on normal 2D displays, the spherical images are projected onto the 2D plane with a limited FOV.
As the FOV becomes wider, the projected image includes more contents and it becomes more similar to human perception, however, the distortion of the image becomes larger, and vice versa.
Therefore, it is required to minimize the subjective distortion of image contents while approaching the possible widest FOV. 



In the last decade, wide-angle projection has been studied in computer vision and graphics community.
Most of the previous works such as rectilinear, stereographic, Pannini projection~\cite{sharpless2010pannini} are based on a single fixed projection model.
Naturally, it has limitation for preserving important contents in the image.
Furthermore, while the center of the projection is less distorted, border regions of an image become more distorted.
To handle this problem, Carroll \textit{et al.}~\cite{carroll2009optimizing} proposed to minimize the distortion of contents such as salient lines and regions through optimization techniques.
%
However, it requires manual extraction of lines to be preserved and is time-consuming.
Rectangling stereographic~\cite{chang2013rectangling} is another contents-preserving projection with automatic line extraction.
However, it does not consider objects of interest in an image, and hard constraint on linear structures can cause large distortion on salient contents.

In this paper, we propose a fast and fully-automated contents-preserving projection method for 360$^\circ$ videos.
We exploit the Pannini projection model~\cite{sharpless2010pannini} as a baseline among many projection models, which has an advantage of preserving not only conformality but also vertical lines.
In addition, we can easily control the behavior of the Pannini projection by adjusting two parameters.
We take the linear structures (\ie lines) and salient regions in images and videos into account for contents-preserving projection which are very important to increase the subjective quality of projected images.
To preserve image contents \emph{locally} and \emph{globally}, we locally apply multiple Pannini projection models with different parameters to a single image in our framework where multiple parameters are adaptively optimized and spatially interpolated based on the image contents.
%
Finally, we consider the temporal consistency of projection to generate temporally consistent and comfortable normal-view videos even under the severe  viewpoint changes.

%
%

The main contributions of this paper are summarized as follows.
First, we propose a contents-preserving projection method optimizing a single Pannini projection model. 
Second, we utilize multiple Pannini projection models to globally minimize contents distortion.
Third, we enforce the temporal consistency for image  projection to produce temporally stable normal-view videos.

The rest of this paper is organized as follows. 
In Sec.~\ref{sec:related_works}, we review various wide-angle projection methods. 
In Sec.~\ref{sec:proposed_method}, we describe the proposed methods to project the spherical image with less distortion, then show experiments on various 360 images and videos in Sec.~\ref{experiments}.
We finally conclude the paper in Sec.~\ref{sec:conclusion}.


\section{Related Works}
\label{sec:related_works}

Spherical image projection methods for wide FOV can be categorized into single-model based, multi-model based, and non-model based methods according to the number of models used for projection as follows.

\textit{Single-model based methods} use a geometric model to project spherical panorama images onto an image plane.
Rectilinear, Stereographic, and Pannini projection~\cite{sharpless2010pannini} models belong to this category. 
Rectilinear projection is the perspective projection from the center point of the spherical image onto the image plane.
This model preserves all lines, but the contents on the margin of a projected image can be extremely stretched and distorted when the FOV is large.
On the other hand, stereographic projection is the perspective projection from the opposite point to the point of tangency as the center of projection. 
This model preserves a conformality of the contents, but not lines.
Pannini projection based on the cylindrical projection keeps the vertical lines straight.
Furthermore, the horizontal lines or radial lines are selectively preserved by the vertical compression. 
The methods mentioned above are very simple and fast, but have a common drawback --- they can not preserve all lines and salient objects simultaneously.

\textit{Multi-model based methods} project images with partially different multiple models depending on the contents such as lines and objects. 
They are comparatively simple and produce less distortion compared with the single-model based projection.
However, they also yield strong distortion at the border of regions where different models are applied.
Zelnik-Manor \textit{et al.}~\cite{zelnik2005squaring} proposed a multi-model based method that applies locally different projections depending on scene structure in the panoramic images with user interaction.
Rectangling stereographic~\cite{chang2013rectangling} projects the spherical image onto a swung surface that is a combination of two orthogonal cylindrical projections with rounded edges.
When the image is divided into four triangular regions with two diagonal lines, it respectively preserves vertical lines in left and right triangular regions and horizontal lines in upper and lower triangular regions of an image.
However, it highly distorts the linear structures and objects straddling the diagonal lines.

\textit{Non-model based methods} try to minimize projection distortion using optimization techniques~\cite{carroll2009optimizing}.
Carroll \textit{et al.}~\cite{carroll2009optimizing} proposed contents-preserving optimization-based projection method.
It produces less distorted images than other approaches and well preserves important contents in the image.
However, it is computationally much more expensive because of the iterative optimization process for every single point. 
Moreover, it requires non-trivial user interaction specifying straight lines to be preserved.

 
\section{Proposed Framework}
\label{sec:proposed_method}

In this section, we present a 
fully-automated contents-aware projection from the 360$^\circ$ videos illustrated in Fig.~\ref{fig:framework}. 
%
%
%
Here, note that, although the proposed framework includes the contents analysis step, we mainly focus on the projection step. The contents analysis steps such as viewpoint selection and salient region extraction can be replaced with any other methods. 
With given viewpoint, we project some portion of a spherical image onto the 2D image plane with less distortion.
We exploit the Pannini projection model~\cite{sharpless2010pannini} as a baseline which has an advantage of preserving not only conformality but also vertical lines well.
As in
~\cite{Zorin1995projection}, we consider two properties, conformality of salient objects and curvature of linear structures, to measure distortions. 

We first extract line segments and salient objects to be preserved automatically. 
Then, we define distortion measures for the Pannini projection model~\cite{sharpless2010pannini} and optimize the Pannini parameters to minimize defined  distortions. If we have multiple salient objects in an image, we compute multiple optimal Pannini projection parameters for multiple salient objects, respectively. 
Afterwards, we perform the model interpolation to minimize contents distortion globally and locally. Since the optimization is performed for a few number of model parameters (two for Pannini projection), our method is much faster than other optimization-based methods. Furthermore, we also consider the temporal consistency of the projection to generate temporally consistent and comfortable perspective videos even under severe viewpoint changes.


{
\begin{algorithm}
\small
\caption{Proposed Contents-Aware Projection}\label{euclid}
\begin{algorithmic}[1]	
\State Input: $I^{s}_{1:N}$, 360 spherical image sequence.
\State Output: $I^{n}_{1:N}$, normal-view image sequence.
 \For{$I^s_t \in I^s_{1:N}$}
        \State Extract $L_t,P_t$ from $I^s_k$ in Sec. 3.1.
		\State Compute $(d^g_t,w^g_t)^-$ using $(L_t,P_t)$ by Eq. (8)
		\State Compute $(d^g_t,w^g_t)^+$ from $((d^g_t,w^g_t)^{-}, (d^g_{t-1},w^g_{t-1})^{+})$ 
        \Statex \hspace{40mm} by Eq. (9)
 \For{$p_i \in P_t$}
 \State Get $L^i_n, P^i_n,$ subsets of $L_t$ and $P_t$ around $p_i$    
 \State Compute $(d^i_t,w^i_t)^-$ using $(L^i_n,P^i_n)$.
\State Compute $(d^i_t,w^i_t)^+$ from 
\Statex \hspace{20mm} $((d^i_t,w^i_t)^{-}, (d^i_{t-1},w^i_{t-1})^{+})$.
 \EndFor
 \State Compute $Proj^{-}_t$ from $(d^g_t,w^g_t,d^i_t,w^i_t, P_t)$ by Eq. (6-7).
 \State Compute $Proj_t^{+}$ from $(Proj_{t}^{-}, Proj_{t-1}^{+})$ by Eq. (10).
 \State Get $I^n_t$ from $I^s_t$ using $Proj_t^{+}.$
 \EndFor
\end{algorithmic}
\end{algorithm}
}

\subsection{Image Content Analysis}
\label{sec:content_analysis}
In this paper, we take the linear structures (\ie lines) and salient regions in images and videos into account for contents-preserving projection which is very important to increase the subjective quality of projected images. Note that the proposed projection method is not dependent on the choice of the image content analysis methods.


To find linear structures in a spherical image, we exploit an advantage of the rectilinear projection which preserves every line in the image.\footnote{Any methods detecting lines on spherical images can be applied.} 
We project a partial spherical image with rectilinear projection and then extract line segments using Line Segment Detector (LSD)~\cite{von2012lsd} from the projected image. 
Each line segment is transformed from image coordinates to spherical coordinates.
In general, distortions of short line segments are less perceivable, so only the line segments longer than a pre-defined threshold are used.

To extract salient objects, we compute scene saliency as the combination of appearance and motion saliency of the image as
\begin{equation}\label{eq:SceneSaliency}
	S_{i}^{scene}=wS_{i}^{appear} + (1-w)S_{i}^{motion},
\end{equation}
where $S_{i}^{scene}$, $S_{i}^{appear}$, and $S_{i}^{motion}$ denote scene saliency map, appearance saliency map, and motion saliency map of the partial image, $i$. 
$\omega$ is a weight parameter.

To find salient objects, we define the appearance saliency as a probability of object existence in the image.
Therefore, we exploit objectness-based object proposals~\cite{alexe2012measuring,BingObj2014,zitnick2014edge} that generate multiple bounding boxes with objectness scores.
The objectness score presents how likely the bounding box contains an object.
We estimate the appearance saliency map by accumulating the objectness score of each object proposal. 
%
To estimate motion saliency $S^{motion}$, we exploit the method proposed in~\cite{hou2007saliency} with optical flow~\cite{weinzaepfel2013deepflow} as an input. 

We assume that objects have higher scene saliency than the background. 
Thus, we extract local peaks as salient objects by applying non-maximum suppression to the scene saliency.
Note that any other saliency detection methods can be applied to our projection method. 

%



\begin{figure*}[th]
\centering 
\includegraphics[width=0.99\linewidth]{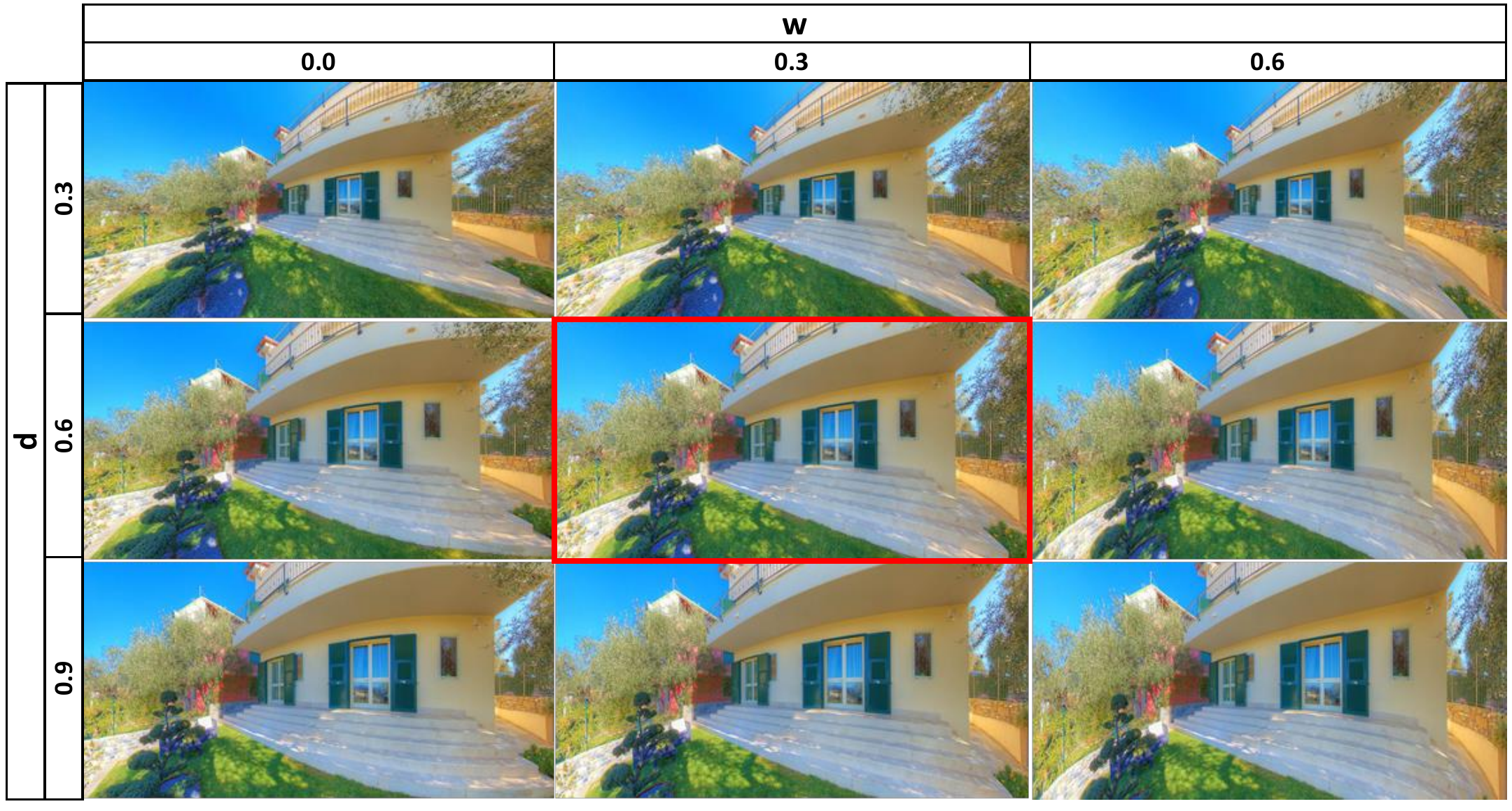}
\caption{\label{fig:optimal_parameter} Estimation of optimal Pannini parameter $d$ and $w$. The red box denotes the Pannini projection result with optimized parameters.}
\vspace{-7pt}
\end{figure*}

\subsection{Optimal Pannini Parameter Estimation}
\label{sec:optimizing_parameter}
To preserve the extracted linear structures and salient points, we use the Pannini projection model as a baseline because it can selectively preserve contents by changing parameters. The Pannini projection model is defined as 
\begin{equation}
\begin{split}
&u_p{ = }\frac { (d+1)\sin  ({ \phi  }_{ p }) }{ d+\cos  ({ \phi  }_{ p }) }, \\
&v_p{ = }\tan  ({ \theta  }_{ p })\left( \frac { (d+1)(1-w) }{ d+\cos  ({ \phi  }_{ p }) } +\frac { w }{ \cos  ({ \phi  }_{ p }) }  \right), 
\end{split}
\end{equation}
where $\theta_p$ and $\phi_p$ denote a point on spherical coordinates, and $u_p$ and $v_p$ denote a correspondence of $\theta_p$ and $\phi_p$ on the image coordinates. 
$d$ and $w$ are control parameters.
$d$ is a distance between the projection plane and the center of projection. 
If $d$ is equal to 0, the projection becomes  rectilinear projection which preserves linear structures but stretches the boundary of perspective images.
If $d$ is equal to 1, it is cylindrical stereographic projection which preserves shape of objects and vertical linear structures but bends radial linear structures.
$w$ is a weighting parameter for vertical compression, which makes horizontal linear structures straight but distorts radial linear structures. Therefore, optimal parameters should be determined depending on contents.


To estimate optimal parameters to preserve both linear structure and salient objects, we define two distortion measures as illustrated in Fig.~\ref{fig:straightness_conformality}. To consider the straightness of linear structure, we define a distortion measure as a distance between the middle point of the line segment and the line which is defined by two endpoints of the line segment on the image plane after projection. 
It is formulated as
{\small
\begin{equation}{ 
\label{line_eq}
D(l)={ \left( \frac { { u }_{ s }({ v }_{ e }-{ v }_{ m })+{ u }_{ e }({ v }_{ m }-{ v }_{ s })+{ u }_{ m }({ v }_{ s }-{ v }_{ e }) }{ \sqrt { { ({ u }_{ s }-{ u }_{ e }) }^{ 2 }+{ ({ v }_{ s }-{ v }_{ e }) }^{ 2 } }  }  \right)  }^{ 2 }},
\end{equation}
}
where subscript $s$, $m$, and $e$ denote the starting, middle, and end point of the line segment, $l$, respectively. If the line is bent when it is projected, the measure has a high value.

To consider shapes of salient objects, we adopt the distortion measure of~\cite{carroll2009optimizing} which is defined as
{
\small	
\begin{equation}{\small
\label{conformality_eq}
C(p)={ \left( cos{ \theta_{p} }\frac {\Delta{u_{p}}} { \Delta{\theta_{p}}} +\frac { {\Delta}{u_{p}} }{ {\Delta}{\phi_{p}}  }  \right)  }^{ 2 }+{ \left( cos{ \theta_{p} }\frac { {\Delta}{v_{p}} }{ {\Delta}{\theta_{p}}  } - \frac { {\Delta}{u_{p}} }{ {\Delta}{\phi_{p}}  }  \right)  }^{ 2 }}.
\end{equation}
}
This measure presents a conformality of a point $p$. With the two measures, we define an objective function as
{
\small	
\begin{equation}
\label{energy}
\begin{split}
E(L_{t},P_{t}) = &\omega _{ d }\sum _{ l\in L_{t} }^{  }{ D(l) } +{ \omega  }_{ c }\sum _{ p\in P_{t} }^{  }{ C(p) },
\end{split}
\end{equation}
}
where $L_t$ is a set of line segments and $P_{t}$ is a set of salient points at frame $t$. 
$\omega_{d}$ and $\omega_{c}$ are weighting parameters that determine which components are more preserved.
The objective function is minimized by the steepest decent method. 
Because it has only two parameters, it is very fast. 
This optimization is globally applied to consider every linear structure and salient object in the image simultaneously. However, it cannot preserve all components simultaneously because it uses a single model to the whole image.
Figure~\ref{fig:optimal_parameter} shows several projection results with various values of the parameters. The parameters, which are obtained by the proposed optimization method, shows the best results.

\subsection{Model Interpolation}
\label{sec:model_interpolation}

To cover remaining distortions, we adopt a multi-model based approach proposed by Zelnik-Manor \textit{et al.}~\cite{zelnik2005squaring}.
The main observation is that the centers of the projected images are less distorted. 
Thus, for one salient point, we project an image around the salient point with a model of which viewpoint is centered at the salient point.
Then, shapes around salient points are preserved. However, regions between salient points have strong distortions because projection models are different from each other. To reduce these distortions, we spatially align multiple models in an image.

First, we set the Pannini projection model with globally optimized parameters as a global model.
The global model determines the whole structure of a perspective image and locations of local models that are projections of salient points.
In this process, if equivalent objects are projected on different locations by the global model and the local models, distortions of these objects should increase in the final results. 
Thus, we applied a transition process to and scaled the local models to match the center of each local model to the global model.
To do this, we define anchor points as the salient points projected by the global model. 
When center points of local models locate on anchor points, shapes projected by local models are aligned with shapes projected by the global model.

After the alignment of the local models, we interpolate local models to fill the regions between salient points smoothly. The interpolated model is defined as  
\begin{equation}\label{model_interpolation_equation}
\begin{aligned}
Proj(u,v)={} & \frac{w_g(u,v)}{Z}Proj_{g}(u,v) \\
                  & + \Sigma_{k}{\frac{w_{A_k}(u,v)}{Z}Proj_{A_k}(u,v)},
\end{aligned}
\end{equation}
\begin{equation}\label{weight_equation}
w_{P}(u,v) = c_{P} e^{-\frac{d_{P}(u,v)^{2}}{2\sigma}}.
\end{equation}
$Proj_{P}$ is the Pannini projection model with $P$ as the center of the model.
It is a backward projection that transforms points on UV coordinates to spherical coordinates.
$Z$ is a normalizing factor. 
$d_P(u,v)$ denotes Euclidean distance between a point $(u,v)$ and a point $P$. 
$g$ and $A_{k}$ represent the center points of the global and the local projection models composing the interpolated projection model. They are subset of $P$ in Eq.~\eqref{weight_equation}. Thus, $Proj_{g}$ represents the global model and $Proj_{A_k}$ represents the local model. 
Therefore, $\{ w_g, \ w_{A_{k}} \} \in w_{P}$ are control parameters to decide the weights of the global and the local models. $w_g, \ w_{A_{k}}$ have their own parameter $c$, that is, $c_g, c_{A_{k}}$.

\begin{figure}[t]
\centering 
\includegraphics[width=1.0\linewidth]{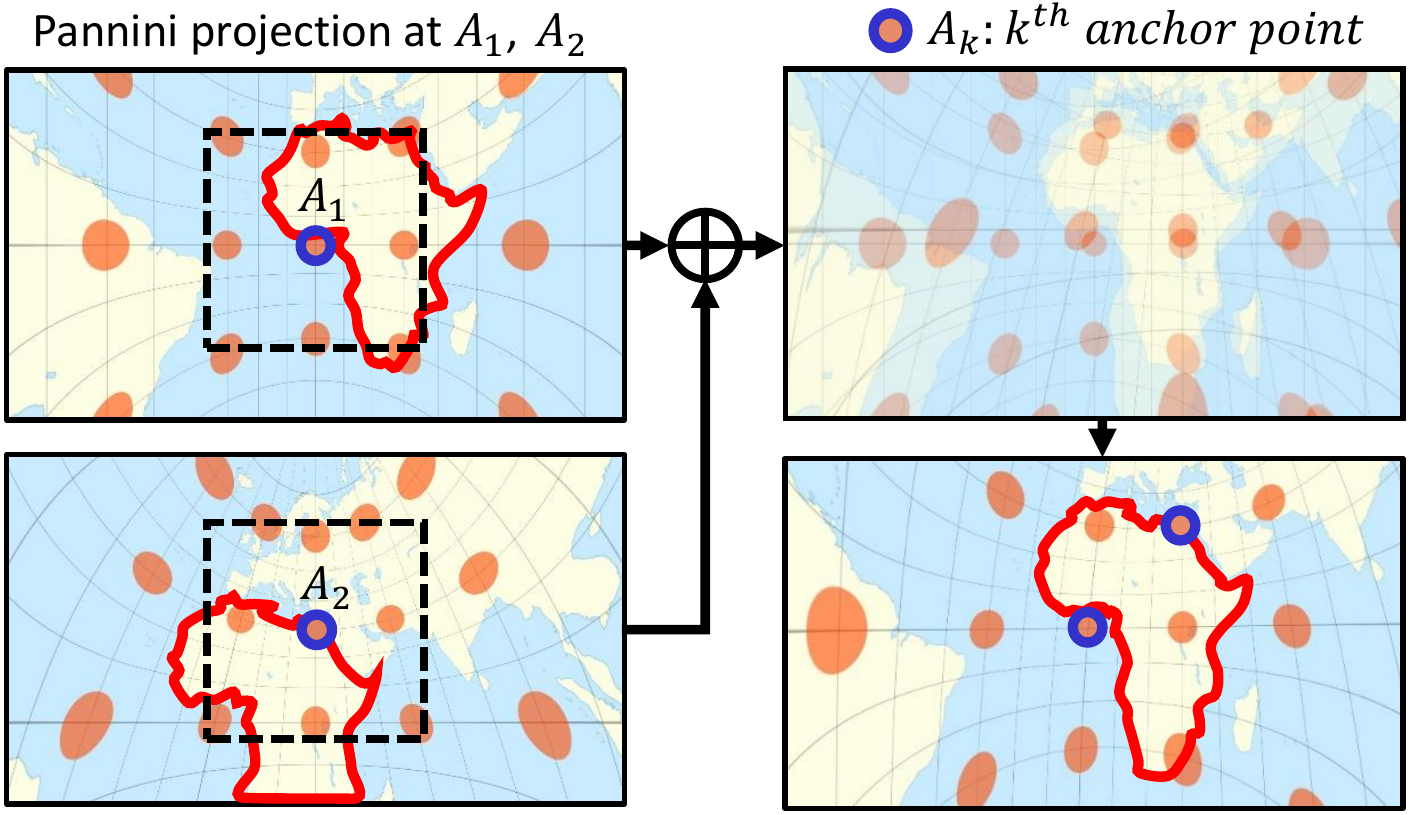}
\caption{\label{fig:model_interpolation}The concept of model interpolation.}
\vspace{-7pt}
\end{figure}

To preserve shapes around salient points, weight $w_{P}(u,v)$ is defined in an exponential form decreasing according to $d_p(u,v)$. 
Then, the region nearby an anchor point $A_k$ is substantially influenced by $Proj_{A_k}$ and projected by $Proj_{A_k}$.
On the other hand, a distant region from $A_k$ is almost unaffected. Therefore, the interpolated model changes smoothly to another projection models. 
Fig.~\ref{fig:model_interpolation} illustrates the concept of the model interpolation. With two anchor points, $A_1$ and $A_2$, $Proj_{A_1}$ and $Proj_{A_2}$ are determined as shown in the left side of the figure. Then, they are aligned and merged to generate the interpolated model.  

\subsection{Temporal Consistency for Video Projection}
The projection model could fluctuate when line segment, salient points, or viewpoints are changed frequently. 
To eliminate the fluctuation for video projection, we enforce the temporal consistency in projection. First, we make parameters of the Pannini projection model consistent temporally. Because parameters determine the behavior of the projection model, a little change of parameters can make severe fluctuation.

To handle this, we add the penalty term $E_{penalty}$ to the objective function for optimization to smooth  parameters as 
%
%
%
\begin{equation}\label{temporal_objective_function}
	\begin{aligned}
		arg \max_{d_k,w_k}&{ \ E(L_t,P_t) + E_{penalty}},\\
		\textrm{where} ~ { E }_{ penalty }={ \omega  }_{ pd }&{ ({ d }_{ t }-d_{ t-1 }) }^{ 2 }+{ \omega  }_{ ps }{ (w_{ t } - w_{ t-1 }) }^{ 2 }.
	\end{aligned}
\end{equation}
Here, $E(L_t,P_t)$ is the objective function in Eq.~\eqref{energy}. $ d_{t}, w_{t}$ indicate the Pannini parameters at frame $t$, and $\omega_{d}$ and $\omega_{s}$ are weighting parameters.
Eq.~\eqref{temporal_objective_function} enforces the estimated parameters in the current frame to be similar with the parameters of the previous frame. 

Furthermore, to make the change of the parameters smoother, we apply the exponential moving average on the parameters estimated from Eq.~\eqref{temporal_objective_function} as    
\begin{equation}\label{parameter_moving_average}
	\begin{aligned}
		d_{t}' &= \omega_{md}d_t+(1-\omega_{md})d_{t-1}',\\    
		w_{t}' &= \omega_{ms}w_t+(1-\omega_{ms})w_{t-1}',
	\end{aligned}
\end{equation}
where $\omega_{md} \in \left[0,1\right]$ and $\omega_{ms} \in \left[0,1\right]$ indicate weighting parameters.

Finally, we use the exponential moving average pixel-wisely to the interpolated model $Proj$ which is defined in Sec.~\ref{sec:model_interpolation}. 
It is defined as   
\begin{equation}\label{model_moving_average}
	\begin{aligned}
		Proj^{t}(u,v) \Leftarrow \omega_{p}Proj^{t}(u,v)+(1-\omega_{p})Proj^{t-1}(u,v),\\    
	\end{aligned}
\end{equation}
where $\omega_{p} \in \left[0,1\right]$ is a weighting parameter. $Proj$ maps a point at $(u, v)$ on the perspective image to the point at $(\phi, \theta)$ on the spherical image. 
Furthermore, we adaptively adjust the weighting parameter $\omega_{p}$ according to change of viewpoint. 
%
%
For example, when the viewpoint remains stationary, inconsistently estimated Pannini projection parameters can cause discomforts to viewers. 
Otherwise, when the viewpoint changes, fluctuations due to inconsistently estimated Pannini projection parameters are less noticeable since contents in the perspective image change rapidly.
Therefore, we increase the weighting parameter $\omega_{p}$ when the viewpoint remains stationary, and we decrease $\omega_{p}$ when the viewpoint changes.  
%
  


\section{Experimental Results}
\label{experiments}

In this section, we compare our projection method with the Pannini projection model~\cite{sharpless2010pannini} which is the baseline of our algorithm, and other state-of-the art projection methods: Carroll \textit{et al.}~\cite{carroll2009optimizing} and Rectangling Stereographic projection~\cite{chang2013rectangling}.  
The 360 $^{\circ}$ image and video datasets for our experiments are collected from the web.

In our experiments, we set $c_g, c_{A_{k}}$, $\omega_{d}$, $\omega_{c}$, $\omega_{pd}$, $\omega_{ps}$, $\omega_{md}$ and $\omega_{ms}$ to 2.0, 1.0, $10^{-3}$, $10^{-4}$, 0.999, $10^{-6}$, $0.9$, and $0.9$, respectively. $\omega_{p}$ is set to $0.99$ when the viewpoints move. Otherwise, $\omega_{s}$ is set to $0.8$. Horizontal FOV and aspect ratio is set to 150$^\circ$ and 16:9 or 170$^\circ$ and 21:9.  



\begin{figure}[t]
	\begin{center}
		\includegraphics[width=1\linewidth]{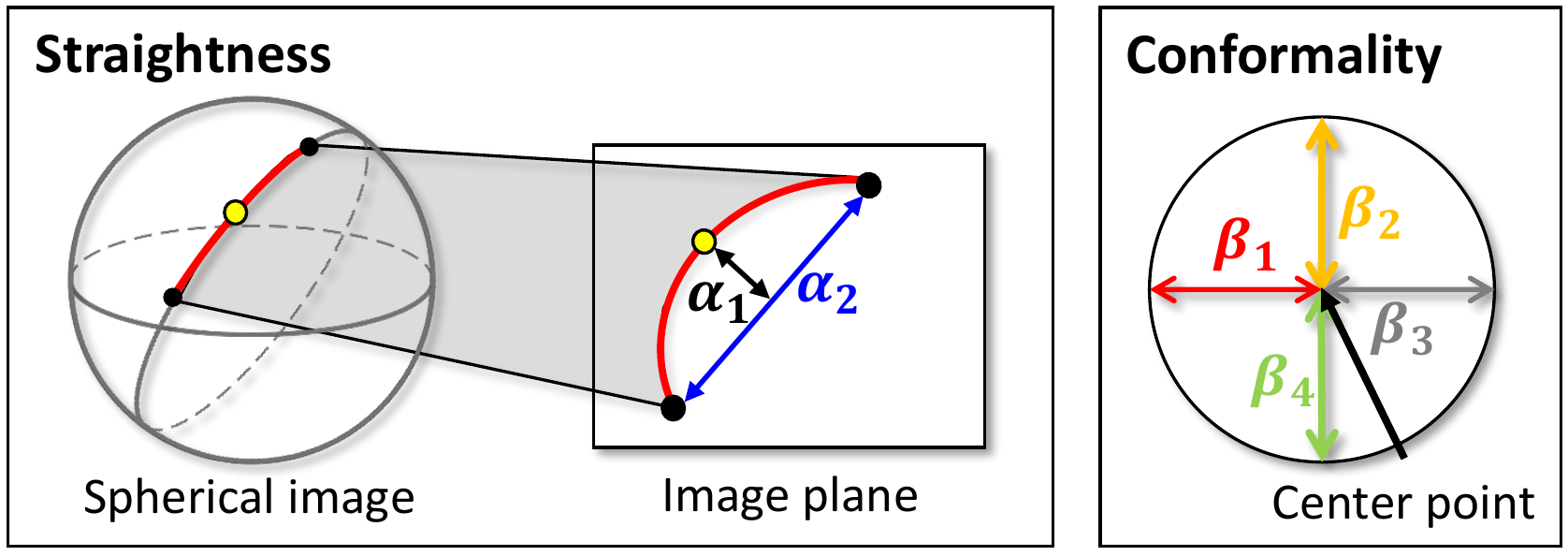}   
	\end{center}
	\vspace{-10pt}
	\caption{Quantitative evaluation measures}
	\label{fig:straightness_conformality}
	\vspace{-12pt}
\end{figure}

\subsection{Quantitative Evaluation}

For quantitative comparisons of the proposed method with other methods, we introduce two distortion measures: straightness and conformality measures. 
The straightness measure indicates the degree in which a line segment in the real-world is bent on the projected image. As shown in Fig.~\ref{fig:straightness_conformality}, given two endpoints of a line segment, we define the straightness measure as 
%
\begin{equation}
\label{eqs:straightness_measure}
Straightness = \frac{\alpha_2}{\alpha_1 + \alpha_2},
\end{equation}
where $\alpha_2$ is the distance between the two endpoints and $\alpha_1$ is the perpendicular distance between the middle point of the curved line segment and a line that joints the two projected endpoints.
The straightness measure is close to 1 if the distortion of the line segment is low, otherwise 0.
As the second measure, we consider the conformality, \ie, measuring the degree to which the appearance of an original spherical image is distorted around salient points. To measure the conformality, we sample four points around a salient point. The points are extracted at the spherical image coordinates moved by 0.1 $rad$ in pitch or roll direction. Then, when the four points are projected on the 2D image plane as in Fig.~\ref{fig:straightness_conformality}, the conformality measure is defined as 
%
\begin{equation}
\label{eqs:conformality_measure}
Conformality = \frac{\min{(\beta_1,\beta_2,\beta_3,\beta_4)}}{\max{(\beta_1,\beta_2,\beta_3,\beta_4)}},
\end{equation}
where $\beta_n$ is a distance between the salient point and the projected sampled point. If the four values from $\beta_1$ to $\beta_4$ are similar to each other, this value is close to 1, which means that the shape around the salient point is less distorted.

\begin{figure}[t]
	\addtolength{\subfigcapskip}{-0.15cm}
	\begin{center}
		\subfigure[Rectilinear]
		{\includegraphics[width=0.48\linewidth]{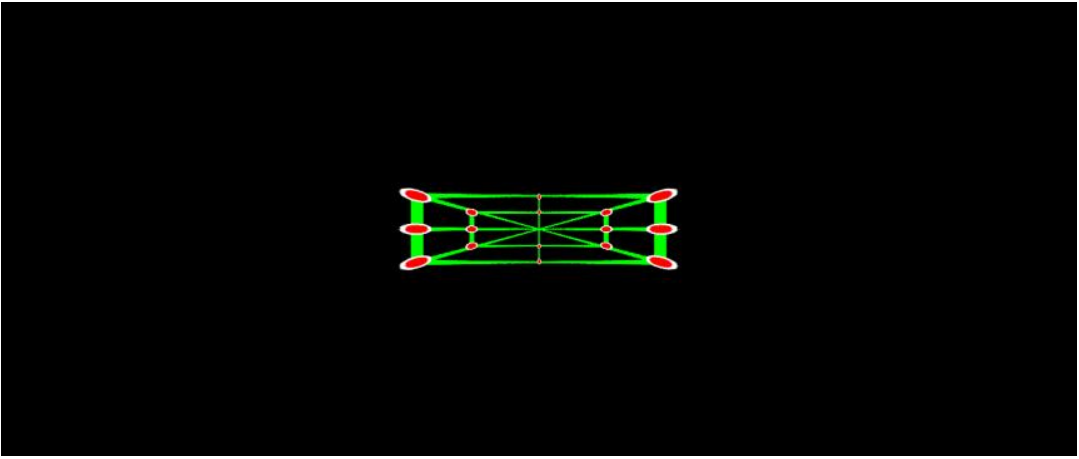}}     
		\subfigure[Pannini \cite{sharpless2010pannini} ($d$=0.5, $w$=0.0)]
		{\includegraphics[width=0.48\linewidth]{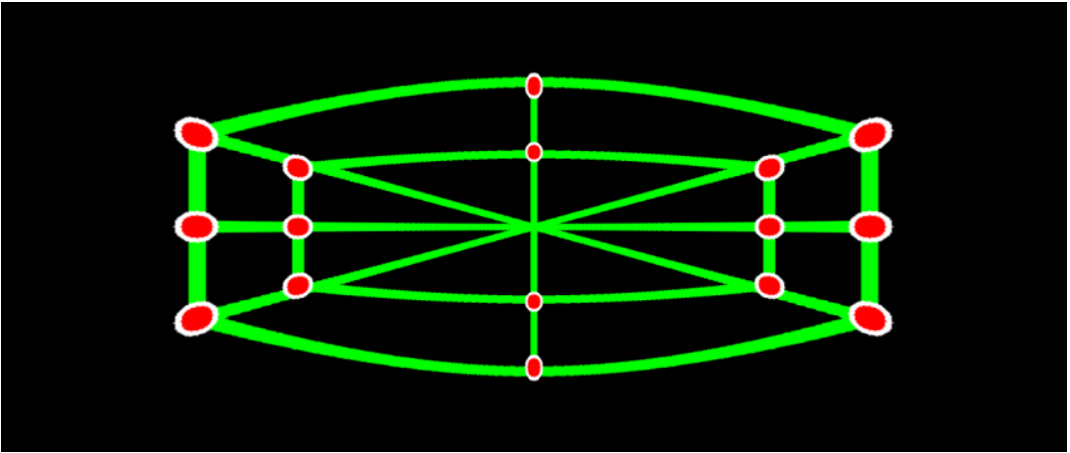}} \\ \vspace{-10pt}
		\subfigure[Pannini \cite{sharpless2010pannini} ($d$=1.0, $w$=0.0)]
		{\includegraphics[width=0.48\linewidth]{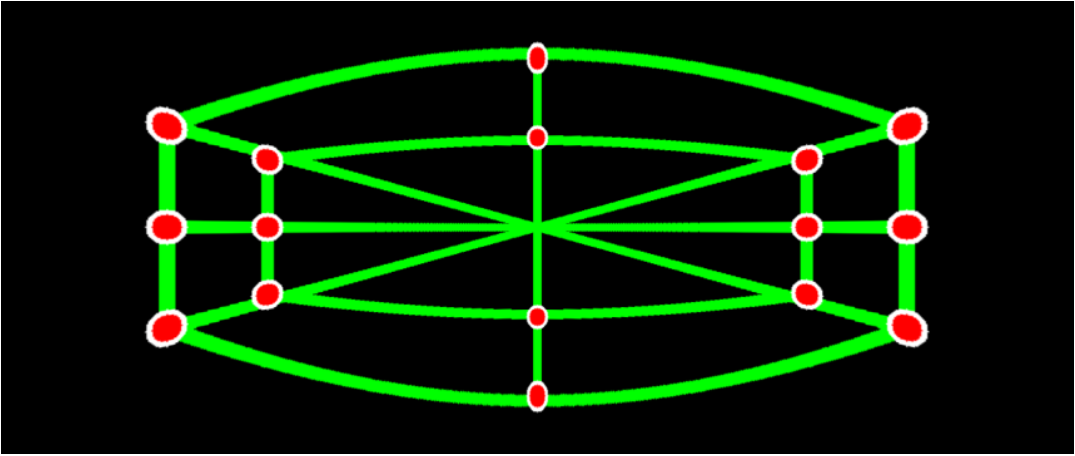}}
		\subfigure[Proposed method]
		{\includegraphics[width=0.48\linewidth]{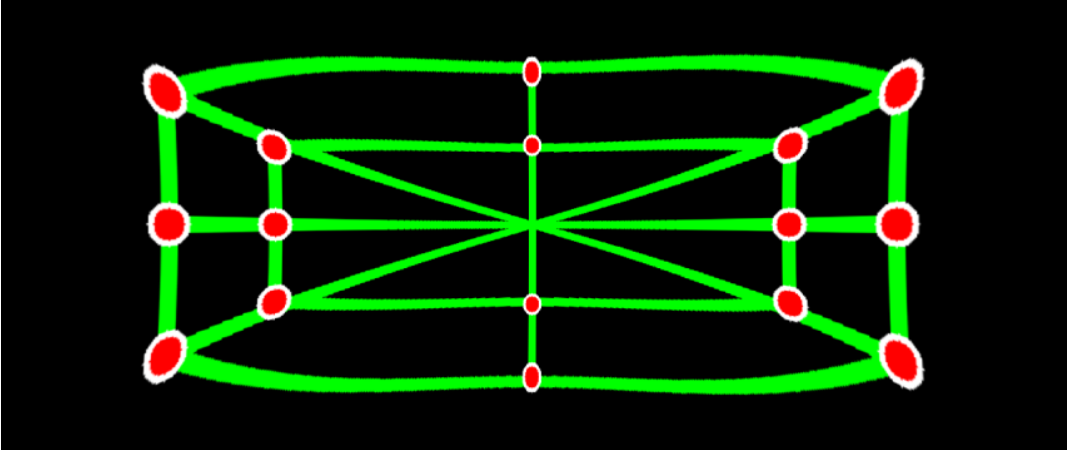}} \\ \vspace{-10pt}
	\end{center}
	\caption{Results of our projection model and other projection models with synthetic input. (FOV = 170$^\circ$, Aspect ratio = 21:9)}
	\label{fig:Quantitative_results170} 
    \vspace{-7pt}
\end{figure}

%
We compared the proposed method with the rectilinear and Pannini projection methods. 
%
To exclude dependency on the content analysis step in Sec.~\ref{sec:content_analysis}, we used manually extracted salient points and line segments as input.
%
%
For quantitative evaluation, we generated synthetic spherical images and then projected them using projection methods with the aspect ratio of 16:9 with the FOV of 150$^\circ$, or with the aspect ration 21:9 with the FOV of 170$^\circ$. 
We used parameters (d=1.0,w=0) for Pannini algorithm, that is commonly used.
Additionally, we included Pannini projection images with d = 0.5  for a variety of comparisons.  
Fig.~\ref{fig:Quantitative_results170} show the results of the projected images, and the red points and green lines represent salient points and line segments, respectively.
Table~\ref{table:quantitative_result_2} represents the results of measuring the straightness and the conformality for each method. The rectilinear projection obtains the highest straightness score but the lowest conformality score, whereas the Pannini method ($d$=0.68) yields high conformality score but low straightness score. Our method achieves high scores in both straightness and conformality. It demonstrates that our method maintains both the straightness and conformality highly in every projection environment. Especially, the minimum straightness scores of our method are higher than other methods except that of the rectilinear method. It means that the proposed method guarantees the competent quality for the straightness compared to the other methods.
%

\begin{table}[t]
	\setlength{\tabcolsep}{8pt}
	\footnotesize
	\caption{Quantitative evaluation of projection results. The best and second best scores in each rank are marked with \textcolor{red}{red} and \textcolor{blue}{blue}. {Straightness and conformality scores are respectively averaged on 11 lines and 16 points.}} \vspace{-3pt}
	\label{table:quantitative_result_2}
	\begin{center}
		\begin{tabular}{c|c|c|c|c}
			\noalign{\hrule height 1pt}    
			\multirow{3}{*}{Algorithm} & \multicolumn{2}{c|}{Straightness}                 														& \multicolumn{2}{c}{Conformality}  																     \\ \cline{2-5}
			& \multicolumn{2}{c|}{Average} 	  									 &  \multicolumn{2}{c}{Average}                                         \\ \cline{2-5} 
			& 16:9      & 21:9    & 16:9       & 21:9               \\ \hline 
			Rectilinear                & {\textcolor{red}{0.9999}} & {\textcolor{red}{0.9999}}  & 0.5235                            & 0.4735                           \\ \hline
			Pannini (d=1.0)           & {0.9639 }         & 0.9747        					    & \BLUE{0.7534}              			  & {\RED{0.8763}}                           \\ \hline
			Pannini (d=0.5)           & 0.9748         					  & 0.9825     			   & {0.7506 } & {0.7298}         \\ \hline
			Optimized Pannini & \BLUE{0.9995} & \RED{{0.9999}}   & 0.5848  & 0.6011 \\ \hline
			Proposed method            & 0.9705         					  & \textcolor{blue}{0.9874}        & {\textcolor{red}{0.7539}}         & \BLUE{0.8407} \\ \noalign{\hrule height 1pt}
		\end{tabular}
	\end{center}
    \vspace{-15pt}
\end{table}

\begin{table*}[h]
	\small
	\centering
	\setlength\tabcolsep{5pt}  
	\caption{Subjective evaluation results. Votes on the 21 sets are averaged, and the range of the average votes is [0, 100]. The proposed methods without and with the model interpolation are denoted as {Optimized Pannini} and {Proposed}, respectively. The best and second best scores are marked with \textcolor{red}{red} and \textcolor{blue}{blue}.} \vspace{3pt}
	\begin{tabular}{c|c|c|c|c|c|c|c|c}
		\noalign{\hrule height 1pt}
		& \multirow{2}{*}{Rectilinear} & \multirow{2}{*}{Stereographic} & \multicolumn{2}{c|}{Pannini~\cite{sharpless2010pannini}}  & \multirow{2}{*}{Carroll~\cite{carroll2009optimizing}} & \multirow{2}{*}{Rect. Stereographic~\cite{chang2013rectangling}} & \multirow{2}{*}{Optimized Pannini} & \multirow{2}{*}{Proposed} \\  \cline{4-5}
		&  &  &  d=1.0 & d=0.5 &  &  &  &   \\ \hline
		Average votes & 19.57 		& 37.24      	& 39.43   & \BLUE{42.76}  &  40.38              & 37.23     & 37.62     & \RED{45.05}  		   \\ \noalign{\hrule height 1pt}
	\end{tabular}
	\label{table:user_study_2}
	\vspace{-9pt}
\end{table*}

\subsection{Subjective Evaluation with User Study}
We perform a user study to verify whether the results of the proposed projection method are comfortable or not with help from the crowd sourcing service \textit{Amazon Mechanical Turk}.
%
%
We generated 21 sets with 8 projection methods: rectilinear, stereographic, Pannini, Carroll, rectangling stereographic, the proposed method without and with model interpolation.
With these 21 x 8 images, we performed blind test with 100 subjects. For each set, we showed 8 projection results in a random order, and asked a question, ''Choose the best 3 photos among below 8 photos'', to the 100 subjects. The votes on each projection method over 21 sets were averaged. 
As shown in the Table~\ref{table:user_study_2}, 
the proposed method with model interpolation produces the most preferred results on average. However, Pannini got higher scores than optimized Pannini.
The proposed method has two-step strategy to project still images. The optimized Pannini is the first step of the proposed method. In the first step, we focus on preserving the straightness of lines as possible. Then, preserving conformality is emphasized in model interpolation stage. For this reason, the optimized Pannini preserves straightness well but does not conformality. Unfortunately, excessively preserved straightness can cause discomfort to viewers as shown in the result on rectilinear projection, which completely preserves the linearity of lines in an image.
Fig.~\ref{fig:mturk_results} shows some results which were evaluated.

\begin{table}[t]
\centering
\caption{Computational time of the proposed method.(second per frame)} \vspace{3pt}
\label{table:processing_time}
\footnotesize
\setlength{\tabcolsep}{5pt}
	\begin{tabular}{c|c|c|c|c}
		\noalign{\hrule height 1pt}	
		Salient point & Line & Pannini parameter & Model &  \multirow{ 2}{*}{\textbf{Total}} \\
        extraction & extraction & optimization & interpolation &  \\ \hline
        0.466 & 0.035 & 0.001 & 0.240 & 0.742 \\ \noalign{\hrule height 1pt}	
	\end{tabular}
    \vspace{-7pt}
\end{table}

\subsection{Qualitative Evaluation}

We perform qualitative comparison for still images.
Pannini projection result is generated with default parameters, ($d$ = 1.0, $w$ = 0.0),  preserving the conformality as much as possible. The result of Carroll \textit{et al.}~\cite{carroll2009optimizing} is obtained with automatically extracted line segments (the same as ours) as inputs. For the proposed method, automatically extracted line segments and salient points are also used as inputs.

\begin{figure*}[ph]
\centering 
\includegraphics[width=0.95\linewidth]{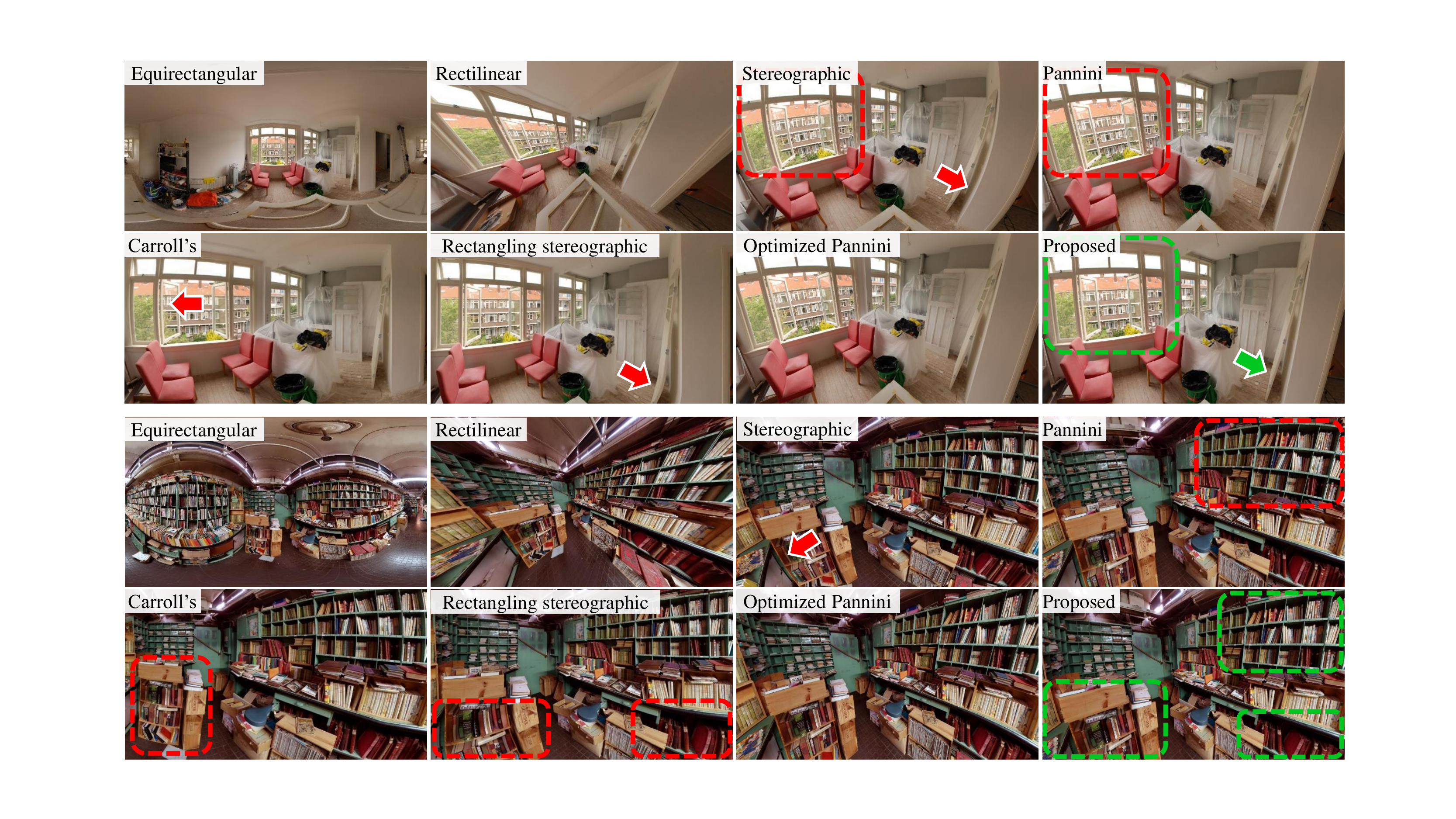}
\caption{\label{fig:mturk_results} Some results evaluated on \textit{Amazon Mechanical Turk}} 
\end{figure*}

{As shwon in Fig.~\ref{fig:mturk_results}, the Pannini projection~\cite{sharpless2010pannini} bends horizontal lines, whereas our results with model interpolation preserves both lines and objects. Our result with only parameter optimization shows that linearity and conformality are more preserved than Pannini. The result of the rectangling stereographic projection looks similar with our result with model interpolation, but lines at the boundaries of the image are bent severely because they are on the border of the different models. Carroll  \textit{et al.}~\cite{carroll2009optimizing} do not preserve the linearity at some lines and end of the lines because the extracted lines are fragmented and missed.}  
More qualitative comparison results are included in our supplementary material.  

\subsection{Results for 360$^\circ$ Videos}
\begin{figure*}[ph]
  \centering
  \includegraphics[width=0.95\linewidth]{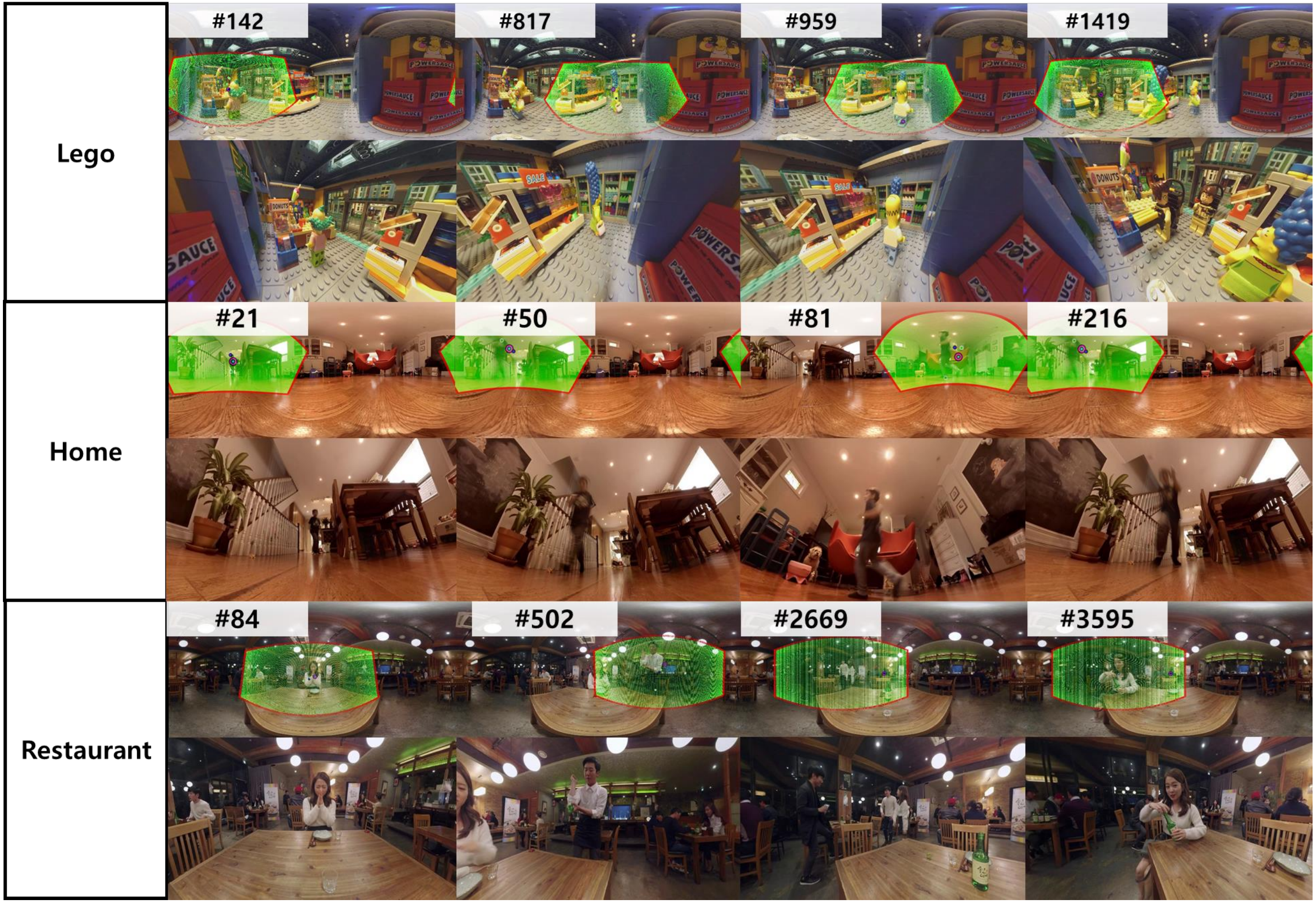}
  \caption{\label{fig:integrated_result} Results of the proposed method for three 360$^\circ$ videos. The proposed method produces temporally consistent and content-preserving projection results from the 360$^\circ$ videos with the given viewpoints.} \vspace{-6pt}
\end{figure*}

We test our method with three 360$^\circ$ videos. Spherical image sequences and viewpoint trajectories are provided as inputs. The results of the proposed method are shown in Fig.~\ref{fig:integrated_result} where the top and bottom images for each dataset represent the spherical image and the projection image, respectively. Green regions in the upper images denote the projected area. We observe that the proposed projection method reduces the distortion around the contents such as straight indoor structures and faces of people. Furthermore, it provides temporally consistent image sequences, which is verified through the video in the supplementary material.

%

\subsection{Computational time}
Table~\ref{table:processing_time} shows computational time for each step of the proposed method with a single core of 4.00GHz. The salient point extraction spends over half of the total computational time in our framework. However, it can be improved by substituting it with faster saliency detection algorithms. Also, the model interpolation takes one-third of the total computational time but it can be reduced by parallelization. The proposed optimization method is much faster than other optimization-based methods. It takes less than 0.001sec (when implemented in C++), whereas Carroll~\cite{carroll2009optimizing} takes about 2sec with GPU (implemented in Photoshop) on the same PC.



\section{Conclusion}
\label{sec:conclusion}
In this paper, we have presented a fully-automated content-aware projection for 360$^\circ$ videos.
To this end, we proposed multi-model based Pannini projection optimization method that preserves both linear structures and salient objects which are automatically extracted.
Additionally, we considered temporal consistency to generate temporally stable videos. 
Experiments including user study show that the proposed projection method is much faster and produces better results than previous content-preserving methods on various environments.

{\flushleft{\bf Acknowledgment.}} {\small 
This work was supported by the Visual Display Business (RAK0117ZZ-21RF) and the Samsung Research Funding Center (SRFC-TC1603-05) of Samsung Electronics.
}


%
{\small
\bibliographystyle{ieee}
\bibliography{egbib}
}
\end{document}